\documentstyle[12pt]{article}
\begin{document}
\renewcommand{\a}{\alpha}
\renewcommand{\b}{\beta}
\renewcommand{\c}{\gamma}
\renewcommand{\d}{\delta}
\newcommand{\th}{\theta}
\newcommand{\TH}{\Theta}
\newcommand{\pa}{\partial}
\newcommand{\g}{\gamma}
\newcommand{\G}{\Gamma}
\newcommand{\A}{\Alpha}
\newcommand{\B}{\Beta}
\newcommand{\D}{\Delta}
\newcommand{\e}{\epsilon}
\newcommand{\E}{\Epsilon}
\newcommand{\z}{\zeta}
\newcommand{\Z}{\Zeta}
\newcommand{\k}{\kappa}
\newcommand{\K}{\Kappa}
\renewcommand{\l}{\lambda}
\renewcommand{\L}{\Lambda}
\newcommand{\m}{\mu}
\newcommand{\M}{\Mu}
\newcommand{\n}{\nu}
\newcommand{\N}{\Nu}
\newcommand{\x}{\chi}
\newcommand{\X}{\Chi}
\newcommand{\p}{\pi}
\newcommand{\r}{\rho}
\newcommand{\R}{\Rho}
\newcommand{\s}{\sigma}
\renewcommand{\S}{\Sigma}
\renewcommand{\t}{\tau}
\newcommand{\T}{\Tau}
\newcommand{\y}{\upsilon}
\newcommand{\Y}{\upsilon}
\renewcommand{\o}{\omega}
\newcommand{\q}{\theta}
\newcommand{\h}{\eta}
\def\ft#1#2{{\textstyle{{#1}\over{#2}}}}
\def\del{\partial}
\def\modb{\bar \Phi}
\def\mat{C}
\def\matb{\bar C}
\def\mpl{M_{\rm Pl}}

\newcommand{\st}{\scriptstyle}
\newcommand{\sst}{\scriptscriptstyle}
\newcommand{\mco}{\multicolumn}
\newcommand{\epp}{\epsilon^{\prime}}
\newcommand{\vep}{\varepsilon}
\newcommand{\ra}{\rightarrow}
\newcommand{\ab}{\bar{\alpha}}
\def\be{\begin{equation}}
\def\ee{\end{equation}}
\def\bea{\begin{eqnarray}}
\def\eea{\end{eqnarray}}

\renewcommand{\section}[1]{\addtocounter{section}{1}
\vspace{5mm} \par \noindent
  {\bf \thesection . #1}\setcounter{subsection}{0}
  \par
   \vspace{2mm} } 
\newcommand{\sectionsub}[1]{\addtocounter{section}{1}
\vspace{5mm} \par \noindent
  {\bf \thesection . #1}\setcounter{subsection}{0}\par}
\renewcommand{\subsection}[1]{\addtocounter{subsection}{1}
\vspace{2.5mm}\par\noindent {\em \thesubsection . #1}\par
 \vspace{0.5mm} }
\renewcommand{\thebibliography}[1]{ {\vspace{5mm}\par \noindent{\bf 
References}\par \vspace{2mm}} 
\list
 {\arabic{enumi}.}{\settowidth\labelwidth{[#1]}\leftmargin\labelwidth
 \advance\leftmargin\labelsep\addtolength{\topsep}{-4em}
 \usecounter{enumi}}
 \def\newblock{\hskip .11em plus .33em minus .07em}
 \sloppy\clubpenalty4000\widowpenalty4000
 \sfcode`\.=1000\relax \setlength{\itemsep}{-0.4em} }
 
\begin{titlepage}
\begin{flushright} THU-95/31\\
hep-th/9601044
\end{flushright}
\vfill
\begin{center}
{\large\bf APPLICATIONS OF SPECIAL GEOMETRY${}^\dagger$ }   \\
\vskip 7.mm
{B. de Wit }\\
\vskip 0.1cm
{\em Institute for Theoretical Physics} \\
{\em Utrecht University}\\
{\em Princetonplein 5, 3508 TA Utrecht, The Netherlands} \\[5mm]
\end{center}
\vfill
 
\begin{center}
{\bf ABSTRACT}
\end{center}
\begin{quote}
We review characteristic features of $N=2$ supersymmetric vector 
multiplets and discuss symplectic reparametrizations and their 
relevance for monopoles and dyons. We 
close with an analysis of perturbative corrections to 
the low-energy effective action of $N=2$ heterotic superstring 
vacua.  %
\vfill      \hrule width 5.cm
\vskip 2.mm
{\small\small
\noindent $^\dagger$ Invited talk given at the Workshop on 
Strings, Gravity \& Related Phenomena, Trieste, 29-30 June 1995; 
to be published in the proceedings.} 
\end{quote}
\begin{flushleft}  
January 1996
\end{flushleft}
\end{titlepage}

\vspace{4mm} 
\begin{center}
{\bf APPLICATIONS OF SPECIAL GEOMETRY }
\vspace{1.4cm}
 
B.~DE~WIT \\
{\em Institute for Theoretical Physics, Utrecht University} \\
{\em Princetonplein 5, 3508 TA Utrecht, The Netherlands} \\
\end{center}             
\centerline{\small ABSTRACT}
\vspace{- 2 mm}  
\begin{quote}\small
We review characteristic features of $N=2$ supersymmetric vector 
multiplets and discuss symplectic reparametrizations and their 
relevance for monopoles and dyons. We 
close with an analysis of perturbative corrections to 
the low-energy effective action of $N=2$ heterotic superstring 
vacua.  %
\end{quote}
\addtocounter{section}{1}
\par \noindent
  {\bf \thesection . Introduction}
  \par
   \vspace{2mm} 

\noindent
Special geometry refers to the target-space geometry of $N=2$ 
supersymmetric vector multiplets, possibly 
coupled to supergravity \cite{DWVP}. The physical states of a vector 
multiplet are described by gauge fields $W^I_\mu$, doublets of 
Majorana spinors $\Omega^I_i$ and complex scalars $X^I$. The 
kinetic term for the scalars is a nonlinear sigma model which 
defines the metric of the target space, the space parametrized by 
the scalar fields. 

The characteristic features of special geometry are as follows. The 
Lagrangian is encoded in a holomorphic prepotential $F(X)$. In rigid 
supersymmetry the fields $X^I$ can be regarded as independent 
coordinates ($I=A=1,\ldots,n$). In the local case there is one 
extra vector multiplet labeled by $I=0$, which provides the 
graviphoton, but now the $n+1$ fields $X^I$ are parametrized in terms 
of $n$ holomorphic coordinates $z^A$. Here one often makes use of
so-called {\it special} coordinates defined by $z^A= X^A/X^0$. 
The target space is K\"ahlerian and the K\"ahler potential is given 
by (the subscripts on $F$ denote differentiation) 
\begin{eqnarray}
K(X,\bar X) &=& -i \bar X^AF_A(X) +i X^A \bar F_A(\bar X)\,,
\nonumber \\
K(z,\bar z) &=& -\log \big[-i \bar X^I(\bar z)F_I(z) +i X^I(z) \bar 
F_I(\bar z)\big]\,,  
\end{eqnarray}
for rigid and for local supersymmetry, respectively. In the latter 
case, the $2n+2$ quantities $(X^I,F_I)$ are parametrized by $n$ 
complex coordinates $z^A$. In a more mathematical context they 
are regarded as holomorphic sections, defined projectively, of a flat 
$Sp(2n+2,{\bf R})$ vector bundle \cite{special,dauria}. The 
ensueing metric satisfies the following curvature relations
\begin{eqnarray}
R^A{}_{\!BC}{}^{\!D} &=& - {\cal W}_{BCE}\bar {\cal W}^{EAD}\,, \nonumber \\
R^A{}_{\!BC}{}^{\!D} &=& 2\delta^A_{(B}\delta^D_{C)} - e^{2K} 
{\cal W}_{BCE}\bar {\cal W}^{EAD}\,, 
\end{eqnarray}
respectively, for the two cases. Here the tensor ${\cal W}_{ABC}$ 
is related to the third derivative of $F(X)$; for global supersymmetry 
one has ${\cal W}_{ABC}= iF_{ABC}$,  while for local supersymmetry 
the relevant expression reads
\begin{equation}
{\cal W}_{ABC} =  iF_{IJK}\big(X(z)\big) \;{\partial
  X^I(z)\over \partial z^A}  {\partial X^J(z)\over \partial z^B}
{\partial X^K(z)\over \partial z^C} \,.
\end{equation}

The kinetic terms for the super-Yang-Mills Lagrangian take the form
\begin{eqnarray}
{\cal L} &=& {\textstyle{i\over 4\pi}}\Big( D_\mu F_I \,D^\mu \bar X^I - 
D_\mu X^I \,D^\mu \bar F_I\Big) - \ft1{8\pi} \,{\rm Im}\,F_{IJ} 
\,\big(\bar\Omega^{iI}  
\buildrel\leftrightarrow\over{D\!\llap/}\, \Omega_i^{I}\big) \label{vlagr}\\
&& -{\textstyle{i\over 16\pi}}
\Big( {\cal N}_{IJ}\,F_{\mu\nu}^{+I}F^{+\mu\nu J}\
-\ \bar{\cal N}_{IJ}\,F_{\mu\nu}^{-I} F^{-\mu\nu J} 
\Big)\,, \nonumber 
\nonumber
\end{eqnarray}
where $\Omega_i$ and $\Omega^i$ are the chiral components of a Majorana
spinor doublet, $F^{\pm I}_{\m\n}$ denote the selfdual and
anti-selfdual field-strength components, and\footnote{%
  In the rigid case, $\cal N$ consists of only the first term and 
  the $I=0$ component is suppressed. 
  In general, $\cal N$ is complex. Its imaginary part 
  is related to the gauge coupling constant, its real part to a 
  generalization of the $\theta$ angle.}
\begin{equation}
{\cal N}_{IJ}=\bar
F_{IJ}+2i {{\rm Im}(F_{IK})\,{\rm Im}(F_{JL})\,X^KX^L\over {\rm
Im}(F_{KL})\,X^KX^L} \,.
\label{Ndef}
\end{equation}
Other terms of the Lagrangian are (we suppress the auxiliary 
fields which may contribute in the presence of hypermultiplets),
\bea
&&-  {\textstyle{i\over16\pi}} \Big(\bar\Omega_i^{I}(D\!\llap/\,
F_{IJ})\Omega^{iJ}   
+\bar\Omega_i^{I}(D\!\llap/\,\bar F_{IJ})\Omega^{iJ}  \Big) \\
&&+{\textstyle{i\over 32\pi}}\Big(F_{IJK}\,
\bar\Omega_i^{I} \sigma\cdot F^{-J} \Omega_j^{K}\varepsilon^{ij}  
-\bar F_{IJK}\, \bar\Omega^{iI} 
\sigma\cdot F^{+J} \Omega^{jK}\varepsilon_{ij}  
 \Big) \nonumber\\
&&+\ft1{2\pi} {\rm Im}\,F_{IJ} \Big( -{\textstyle{1\over 2}}g \,f^J_{KL} \big[\bar\Omega{}^I_i\bar 
X{}^K\Omega^L_j \varepsilon^{ij} +\bar\Omega{}^{iI}  
X{}^K\Omega^{jL} \varepsilon_{ij}\big] \nonumber \\
&&\hskip 23mm 
+g^2 (f^I_{KL}\bar X^KX^L) (f^J_{MN}\bar 
X^MX^N)\Big)\,,\nonumber 
\eea
where the $f^I_{JK}$ are the structure constants of the gauge 
group.

Special geometry is relevant for the moduli of Calabi-Yau 
three-folds. This intriguing connection can be  
understood in the context of type-II superstrings, whose 
compactification on Calabi-Yau manifolds leads to 
four-dimensional effective actions with local $N=2$ 
supersymmetry. In these field theories the sigma-model fields have no 
potential. Therefore their vacuum-expectation values are 
undetermined and parametrize the (classical) ground states 
of this field theory, up to certain equivalence transformations. 
The metric of   
the non-linear sigma model (in four space-time dimensions) is 
therefore related to the metric on the moduli space 
of superstring ground states associated with Calabi-Yau 
spaces \cite{Seiberg}. Therefore this moduli space must exhibit 
special geometry. For the complex-structure moduli, the periods 
of the $(3,0)$ form correspond to the sections $(X^I,F_J)$, which 
transform under symplectic rotations induced by changes in the 
corresponding homology basis \cite{special,Cand}. These 
symplectic transformations exist already at the level of the 
supersymmetric field theory, as we discuss below.

\section{Symplectic reparametrizations}
\noindent{}From the Lagrangian (\ref{vlagr}) one defines the tensors 
\be
G^+_{\mu\nu I}={\cal N}_{IJ}F^{+J}_{\mu\nu}\,,\quad G^-_{\mu\nu
I}=\bar{\cal N}_{IJ}F^{-J}_{\mu\nu}\,, \label{defG}
\ee
so that the Bianchi identities and equations of motion
for the Abelian gauge fields can be
written as
\begin{equation}
\partial^\mu \big(F^{+I}_{\m\n} -F^{-I}_{\m\n}\big)
=0\,,\qquad
\partial^\mu \big(G_{\mu\nu I}^+ -G^-_{\m\n I}\big) =0\,.
\label{Maxwell}
\end{equation}
These are invariant under the transformation
\be
\pmatrix{F^{+I}_{\mu\nu}\cr  G^+_{\mu\nu I}\cr} \longrightarrow 
\pmatrix{U&Z\cr W&V\cr} \pmatrix{F^{+I}_{\mu\nu}\cr  G^+_{\mu\nu 
I}\cr} \,,\label{FGdual}
\ee
where $U^I_{\,J}$, $V_I^{\,J}$, $W_{IJ}$ and $Z^{IJ}$ are 
constant real  $(n+1)\times(n+1)$ submatrices.
The transformations for the anti-selfdual tensors follow by
complex conjugation. From (\ref{defG}) and (\ref{FGdual}) one derives 
that $\cal N$ transforms as
\begin{equation}
{\cal N}_{IJ} \longrightarrow (V_I{}^K {\cal N}_{KL}+ W_{IL} )\,
\big[(U+ Z{\cal N})^{-1}\big]^L{}_J  \,.\label{nchange}
\end{equation}
To ensure that $\cal N$ remains a symmetric tensor, at 
least in the generic case, the transformation  
(\ref{FGdual}) must be an element of $Sp(2n+2,{\bf R})$  
(we disregard a uniform scale transformation). 
The required change of $\cal N$ is induced by a change of the scalar 
fields, implied by
\be
\pmatrix{X^{I}\cr  F_{I}\cr} \longrightarrow  \pmatrix{\tilde 
X^I\cr\tilde F_I\cr}=
\pmatrix{U&Z\cr W&V\cr} \pmatrix{X^{I}\cr  F_I\cr}\,. 
\label{transX}
\ee
In this transformation we include a change of $F_I$. Because the 
transformation belongs to $Sp(2n+2,{\bf R})$, one can show that 
the new quantities $\tilde F_I$ can be written as  
the derivatives of a new function $\tilde F(\tilde X)$. 
The new but equivalent set of equations of motion one obtains by 
means of the symplectic transformation (properly extended to other 
fields), follows from the Lagrangian based on $\tilde F$. In 
special cases $F$ remains unchanged, $\tilde F(\tilde X) = F(\tilde 
X)$, so that the theory is {\it invariant} under the 
corresponding transformations. 

The symplectic transformations act on the (anti-)selfdual 
components of the field strengths (cf. (\ref{FGdual})). Such 
duality transformations are known to appear in all extended  
supergravity theories in four space-time dimensions  \cite{dual}
and were studied extensively as continuous invariances of the 
equations of motions (for $N=2$, we refer to \cite{DWVP,BEC,
dWVPVan}). For instance, the class of functions
\begin{equation}
F(X)= d_{ABC}{X^AX^BX^C\over X^0}\,,\label{dfunction}
\end{equation}
generally leads to symplectic invariances associated with 
finite parameters $b^A$ and defined by 
\begin{eqnarray}
U(b)= V^{\rm T}(-b)=  \pmatrix{1&0\cr
\noalign{\vskip 2mm}
               b^A&{\bf 1}_n\cr} \,,\quad
W(b)= \pmatrix{-(d\,bbb)& - 3(d\,bb)_B\cr
\noalign{\vskip 3mm}
               3(d\,bb)_A& 6 (d\,b)_{AB}\cr} \,,\quad
Z(b)=0\,. \label{btransform}
\end{eqnarray}
In terms of the special coordinates, these give rise to
\begin{equation}
z^A\longrightarrow z^A + b^A  \,.
\end{equation}
Symplectic transformations with $Z=0$ can always be realized on 
the vector potentials and thus leave the Lagrangian invariant 
(possibly up to a total  
divergence corresponding to a shift in the $\theta$ angles).

Later it was realized that symplectic transformations can be 
used to relate different functions $F(X)$  describing 
equivalent equations of motion \cite{CecFerGir}. These 
reparametrizations were exploited in the context of Calabi-Yau 
manifolds and, more recently, for 
$N=2$ nonabelian gauge theories to describe singularities in the 
Wilsonian action that originate from the emergence of massless 
states at strong coupling in terms of a dual theory at weak 
coupling \cite{SW}. Generically duality transformations mix 
electric and magnetic fields.   The  
interchange of electric and magnetic fields is known as    
electric-magnetic duality. For instance, for $U=V=0$ and $W=-Z ={\bf 1}$, 
$F^{+I}_{\mu\nu}$ and   
$G^+_{\mu\nu I}$ are simply interchanged, while $\cal N$ 
transforms into  $-{\cal N}^{-1}$. Since the coupling constants 
are thus replaced by their inverses, electric-magnetic duality 
relates the strong- and  
weak-coupling description of the theory. Electric-magnetic duality is 
a special case of so-called $S$ duality. The coupling constant 
inversion is  part of an $SL(2,{\bf Z})$ group. This duality 
is also known in the context of string theory \cite{FLQ} and  
lattice gauge theories \cite{Cardy}. 
Other symplectic transformations induce a  
shift of the generalized $\theta$ angles. In nonabelian  
gauge theories $\theta$ is periodic, so that $\cal N$ is 
defined up to the addition of certain {\it discrete} real 
constants.  This is a generic feature, both here as well as in 
string theory; the nonperturbative  dynamics restricts the 
symplectic transformations to a discrete subgroup.   For an 
earlier account of confinement phases in nonabelian gauge 
theories where duality transformations were important, see 
\cite{thooft}.  

\section{Semiclassical consequences of monopoles and dyons}
\noindent 
To elucidate some important features of  
the symplectic reparametrizations, let us discuss the effective 
action of abelian gauge fields, possibly obtained from a 
nonabelian theory by integrating out certain fields. We 
write the matrix $\cal N$ in terms of generalized coupling 
constants and $\theta$ angles, according to
\be
{\cal N}_{IJ}= {\theta_{IJ}\over 2\pi} -i {4\pi\over g^2_{IJ}}  
\,.\label{caln}      
\ee
This matrix can be viewed as a generalization of the 
permeability and permittivity that is conventionally used in the 
treatment of electromagnetic fields in the presence of a medium. 
The fields  
$G_{\mu\nu I}$ are thus generalizations of the displacement and 
magnetic fields, while $F^I_{\mu\nu}$ corresponds to the electric 
fields and magnetic inductions. So far we have considered an abelian 
theory without charges. It is straightforward to introduce 
electric charges by including an electric current in the 
Lagrangian. To consider duality tranformations one must 
also include magnetic currents into the field equations, so 
that when electric fields tranform into magnetic fields and vice 
versa, the electric and magnetic currents transform accordingly. 
These magnetic currents occur as sources in the Bianchi identity 
and describe magnetic monopoles. 

Electric and magnetic charges are conveniently defined in terms 
of flux integrals over closed spatial surfaces that surround the 
charged objects,
\bea
\oint_{\partial V} (F^+ + F^-)^I \!&=& \! 2\pi \,q_{\rm m}^{I}\,,\nonumber \\
\oint_{\partial V} (G^+ + G^-)_I \!&=&  \! -2\pi \,q_{{\rm e}I} \,. 
\label{charges}
\eea
With these definitions a static point charge at the origin 
exhibits magnetic inductions and electric fields equal to $\vec 
r/(4\pi r^3)$ times $2\pi q_{\rm m}^{I}$ and 
$\ft12g^2(q_{{\rm e}I} + q_{\rm m}^J \,\theta_{IJ}/2\pi)$, 
respectively. Note that $q_{\rm e}$ does not 
coincide with the electric charge. The $\theta$-dependent mixing 
of the electric and magnetic charges was first noted in 
\cite{witten} and follows directly from the generalized Maxwell 
equations in the presence of the $\theta$ angle \cite{salam,
thooft}. From (\ref{charges}) it follows that the 
charges must transform under symplectic rotations according to
\be
\pmatrix{q_{\rm m}^{I}\cr  -q_{{\rm e}I}\cr} \longrightarrow 
\pmatrix{U&Z\cr W&V\cr} \pmatrix{q_{\rm m}^{I}\cr  -q_{{\rm 
e}I}\cr} \,. 
\ee
As is well known, the charges are subject to a generalized Dirac 
quantization condition, due to Schwinger and Zwanziger 
\cite{Schwinger}, according to which $q_{\rm  
e}q^\prime_{\rm m}-q_{\rm m}q^\prime_{\rm e}$ must be a multiple 
of $2\hbar$. This implies that the allowed 
electric and magnetic charges comprise a lattice such that surface 
elements spanned by the lattice vectors are equal to a 
multiple of the Dirac unit $2\hbar$. In 
addition, this lattice should be consistent with the 
periodicity of the $\theta$ angle%
\footnote{%
   The normalization of the $\theta$ angle is 
   fixed by the assumption that instantons yield an integer value 
   for the Pontryagin index $(32\pi^2)^{-1}\int{\rm d}^4 x 
   \;{}^\ast\!F F$ in a nonabelian extension of the theory.%
}, $\theta\to \theta + 2\pi$, which corresponds to ${\cal 
N} \to {\cal N} + {\bf 1}$. This shift is associated with a symplectic 
transformation with $U=V=W={\bf 1}$ and $Z=0$, so that the 
charges transform as $q_{\rm e}\to q_{\rm e}- q_{\rm m}$ 
and $q_{\rm m}\to q_{\rm m}$. This transformation must be
contained in the discrete subgroup $Sp(2n+2,{\bf Z})$ that leaves 
the charge lattice invariant.  

As observed by Olive and Witten \cite{olive}, $q_{\rm e}$ and 
$q_{\rm m}$ emerge as surface integrals in the supersymmetry 
algebra. We derive this in a slightly more general setting for a 
supersymmetric Yang-Mills theory based on a holomorphic function 
$F(X)$. In that case the supercurrent reads
\be
J_{\mu i} = \ft1{2\pi} {\rm Im}\,F_{IJ}\Big\{D\!\llap/ \,\bar 
X^I\gamma_\mu\Omega_i^J -  \varepsilon_{ij} \Big[
{\textstyle{1\over 2}}\sigma\cdot F^{-I}- g 
\,f^I_{MN}\bar X^MX^N\Big]\gamma_\mu\Omega^{jJ}\Big\} \,.
\ee
In the abelian limit one can show that this result remains the
same under symplectic reparametrizations. This is not surprising
in view of the fact that the symplectic reparametrizations are
also applicable in a supergravity background \cite{DWVP}. For a
more general discussion of the effect of supergravity and chiral 
backgrounds, see \cite {bdewit}. 

From the Dirac brackets (suppressing explicit spinor indices)
\bea
\big\{\Omega_{j}^I( x),\bar\Omega^{iJ}(y)\big\}_{x_0=y_0} &=& 4\pi\hbar 
\, [({\rm Im}\,F)^{-1}]^{IJ}\, \delta^i_j\,\Big({1+\gamma_5\over 
2}\gamma_4\Big) \;\delta^3(\vec x-\vec y)\,, \nonumber\\
\big\{\Omega^{iI}( x),\bar\Omega_{j}^{J}( y)\big\}_{x_0=y_0} & =& 
4\pi\hbar \, [({\rm Im}\,F)^{-1}]^{IJ}\, \delta^i_j\,
\Big({1-\gamma_5\over  
2}\gamma_4\Big) \;\delta^3(\vec x-\vec y) \,,
\eea
we immediately determine the
anticommutators of the supersymmetry charges $Q_i\equiv  \int
{\rm d}^3x\; J^0_i$, at 
least as far as their bosonic contributions are concerned. The
first one is ( $a,b,c$ label space coordinates)
\be
\{Q_i,\bar Q^j\} = i\hbar\,\d^i_j (1-\gamma_5) \int {\rm d}^3x
\;\Big\{\g_\m\, T^{\mu 0} +\ft1{8\pi}\g_a \,\varepsilon^{abc}  \,
\partial_b\Big(F_I\, \buildrel\leftrightarrow\over D_c \bar X^I 
-X^I\,\buildrel\leftrightarrow\over D_c \bar F_I \Big)\Big\} \,,
\ee
where $T^{\m\n}$ is the energy-momentum tensor (which in the
abelian limit also 
preserves its form under symplectic reparametrizations) and the
total divergence leads to a surface term which can be dropped. 
The second anticommutator can be written as 
\bea
\{Q_i,\bar Q_j\} = -{i\hbar\over 4\pi} (1-\gamma_5)
\varepsilon_{ij} 
\int {\rm d}^3x  \!\!&\Big\{&\!\!\!\!\varepsilon^{abc}\Big[ 
(D_a \bar X^I)  
(G^++  G^-)_{Ibc}  - (D_a \bar F_I) (F^++ F^-)_{bc}^{I}\Big]
\nonumber\\ 
&&+ 4\,{\rm Im}\,F_{IJ} \,(D_0\bar X^I) \,f^J_{KL} \bar X^KX^L 
\Big\} \,.
\eea
Using the equations of motion and (\ref{charges}), this yields 
the following expression for the central charge,
\be
\{Q_i,\bar Q_j\} = i\hbar\, (1-\gamma_5)\,
\varepsilon_{ij} \,\Big\{ \bar X^I\,q_{{\rm e}I}+ \bar 
F_I\, q_{\rm m}^I \Big\}\,,
\ee
where $\bar X^I$ and $\bar F_I$ represent the constant values 
taken by these quantities at spatial infinity. The right-hand 
side manifestly preserves its form under symplectic transformations, 
precisely as expected. This representation of the central charge 
plays an important role in the analysis of \cite{SW}. 

\section{$N=2$ Heterotic vacua}
\noindent
In heterotic string vacua the $N=2$ space-time supersymmetry 
charges reside entirely in the right-moving sector. This sector 
decomposes into a $c=3$ and a $c=6$ superconformal field theory 
with $N=2$ and 4  
world-sheet supersymmetry, respectively. The massless spectrum 
that emerges from this sector together with the four-dimensional 
space-time sector yields the graviton, an antisymmetric tensor, 
the dilaton and two abelian boson fields. Together with 
corresponding fermions, they constitute the supergravity multiplet 
tegether with a so-called vector-tensor multiplet\footnote{%
  This multiplet was first considered in \cite{SSW} in a study of 
  off-shell representations of supersymmetric Yang-Mills theory. 
  The multiplet has an off-shell central charge, which couples to 
  a gauge field in supergravity \cite{CDWFKST}.}  \cite{DWKLL}. 
On shell the tensor field can be converted into a scalar which 
combines with the dilaton into a complex scalar $S$. The 
latter then belongs to an $N=2$ vector multiplet. Other vector 
multiplets originate from the left-moving sector. The simplest 
case corresponds to the toroidal compactifications from 
the six-dimensional $N=1$ heterotic theory. In that case one has 
two corresponding moduli, denoted by $T$ and $U$. The toroidal 
compactifications can be continuously deformed by nontrivial 
Wilson lines in the gauge group associated with the gauge fields 
accompanying the two torus periods; 
those deformations give rise to extra vector multiplets. 
Our arguments below are not restricted to toroidal 
compactifications, however, and we assume an arbitrary number of 
moduli. For definiteness we take this number to exceed 2, so 
that we will be dealing with at least 3 vector multiplets.  

Classically the effective action that describes the vector 
multiplets relevant for $N=2$ heterotic vacua is restricted by 
the fact that the dilaton (the real part of $S$) couples 
universally while its imaginary part acts as a generalized $\theta$ 
angle and must be invariant under constants shifts. This uniquely 
determines the holomorphic homogeneous function \cite{FVP} (up 
to symplectic reparametrizations),
\be
F_{\rm class}(X) = - {X^1\over X^0} \Big[X^2X^3 - 
\sum_{I\geq 4}^n (X^I)^2\Big] \,,  \label{function}
\ee
which corresponds to the product manifold $[SU(1,
1)/U(1)]\times[SO(2,n-1)/(SO(2)\times SO(n-1))]$. The $SU(1,
1)/U(1)$ coordinate is the dilaton field $iS=X^1/X^0$, whose real 
part corresponds the string coupling constant. Other 
moduli are given by $iT=X^2/X^0$, $iU=X^3/X^0$, 
etc.; these moduli transform under the target-space duality  
group $SO(2,n-1)$. 

The objective is to consider the perturbative string corrections 
to (\ref{function}). For these corrections the dilaton 
field $S$ acts as a loop-counting parameter, so that $n$-loop 
corrections will be inversely proportional to $(X^1)^{n-1}$. 
Perturbatively the effective action 
must be invariant under continuous shifts of the imaginary 
part of $S$ (proportional to the $\theta$ angle). At the 
same time the dependence on $S$ in the function $F(X)$ should 
remain holomorphic to all orders in perturbation theory. These
two requirements restrict the 
possible additions to (\ref{function}) to be independent of $X^1$, so 
that there are no perturbative corrections beyond one-loop. In 
addition there  are nonperturbative corrections, 
which are not covered by the above argument as they are only 
invariant under discrete $2\pi$-shifts of the $\theta$ angle.  

The one-loop corrections should preserve the invariance under  
target-space duality. However, a subtlety may occur as we expect 
the corrections to exhibit a certain lack of single-valuedness 
due to the presence of singular points in the  
moduli space where massive string states become massless. One may 
wonder whether there is an appropriate symplectic basis for the 
periods $(X^I,F_I)$ in which to address these questions. The 
basis defined by  
(\ref{function}) has two conspicuous features. First of all, 
the gauge couplings do not all become weak in the large dilaton 
limit, so that this does not seem a good starting point for 
setting up consistent string perturbation theory. Secondly, the 
$SO(2,n-1)$ invariance is realized by duality transformations, so 
that the equations of motion, but not the classical Lagrangian, 
are left invariant. These duality transformations 
involve inversions of the gauge couplings and it is therefore 
plausible that these two features are related. Indeed it is 
possible to redefine the periods by means of a symplectic 
reparametrization, such that all gauge couplings vanish uniformly 
in the large-dilaton limit and, at the same time, the classical 
Lagrangian is strictly invariant under $SO(2,n-1)$. The new 
periods are defined by\footnote{%
   Note that the following results are expressed in 
   terms of the periods and not in terms of a new function $\hat 
   F(\hat X)$.    The reason is that the $\hat X^I$ are not 
   independent and no suitable  function exists. 
   Fortunately the latter is merely a   
   technical problem as the full Lagrangian can still be written 
   down consistently in terms of the periods and their derivatives 
   \cite{Ceresole}.}%
\be
\hat X^I= (X^0, F_1, X^2, \cdots, X^n)\,,\qquad \hat F_I=(F_0, 
-X^1, F_2,\cdots , F_n)\,.
\ee
Now $SO(2,n-1)$ acts linearly on both $\hat X^I$ and $\hat F_I$ 
separately,
\be
\hat X^I \rightarrow   \hat U{}^I_{\,J}\, \hat X^J,\qquad
\hat F_I \rightarrow   \hat V_I{}^J \,\hat F_J \,,
\label{oltrans}
\ee
where $\hat V = (\hat U^{\rm T})^{-1}$ and $\hat U$ is an $SO(2,
n-1)$ matrix. The above transformations pertain to the classical case 
(\ref{function}) and leave the corresponding Lagrangian invariant. 
According to (\ref{oltrans}) $X^0,X^2,\cdots, X^n$ 
transform among themselves, but in a nonlinear fashion. 

Can these transformations be modified when one-loop  
corrections are included and $F(X)$ is no longer single-valued? 
The answer to this question follows from the observation that 
adding a one-loop correction $F_{\rm 1-loop}(X)$ (which is 
$X^1$-independent) to  $F_{\rm class}(X)$ does not affect the 
definitions of the $\hat X^I$. Furthermore the   
transformations of $X^0,X^2,\cdots,X^n$, and therefore those of 
the $\hat X^I$, should remain the same at the 
quantum level, as these fields have fixed relations to their 
string vertex operators. This is not so for $X^1$, which 
is obtained by conversion of a vector-tensor multiplet. On the 
other hand, the target-space duality transformations should still 
take a symplectic form. Therefore the modifications are 
restricted to 
\be
\hat X^I \rightarrow  \hat U{}^I_{\,J}\, \hat X^J,\qquad
\hat F_I \rightarrow   \hat V_I{}^J \,\hat F_J\ +  \hat W_{IJ}\, 
\hat X^J\ , \label{newtrans}
\ee
where $\hat V = (\hat U^{\rm T})^{-1}$ and $\hat W = \hat V 
\Lambda$ with $\Lambda$ a real symmetric matrix. The 
corresponding Lagrangian is now no longer invariant but changes 
by a total derivative proportional to $\Lambda$. 
The latter is induced as a result of monodromies around the 
semi-classical singularities in moduli space. For finite 
string coupling we expect only some discrete subgroup of $SO(2,n-1)$ 
to be relevant \cite{SEN}; $\Lambda$ is then integer-valued as 
well. 

Furthermore one can show that $F_{\rm 1-loop}(X)$ can be written 
as $F_{\rm 1-loop}(X)= {1\over 2} \hat  F_I\hat X^I$. Substitution 
of (\ref{newtrans}) then yields at once the  
variation under target-space duality 
\be
F_{\rm 1-loop}(\tilde X) = F_{\rm 1-loop}(X) +\ft12 \Lambda_{IJ} 
\hat X^I\hat X^J\,. \label{deltaF1}
\ee
The one-loop contribution to the function 
${\cal F}_{\rm 1-loop}\equiv  i(X^0)^{-2}F_{\rm 1-loop}(X)$ must 
therefore be invariant under target-space duality transformations, up to a 
restricted polynomial of the moduli with discrete real 
coefficients \cite{DWKLL}. According to (\ref{deltaF1}) $F_{\rm 
1-loop}$ is multivalued and the ambiguities in this function amount 
to the quadratic polynomial in the variables $\hat X^I$ (to see 
this consider a transformation with $U={\bf 1}$, so that the 
$\hat X^I$ remain unchanged, but $\Lambda\not= 0$). 

As an explicit example one may consider toroidal 
compactifications of six-dimen\-sion\-al  
$N=1$ string vacua, where we have only $T$ and $U$. The 
transformation $T\to (aT-ib)/(icT+d)$ with integer 
parameters satisfying $ad-bc =1$, then induces the following result on 
the one-loop correction (one can also consider a similar 
transformation of $U$),  
\be
{\cal F}_{\rm 1-loop}(T,U)\to (icT+d)^{-2} [{\cal F}_{\rm 
1-loop}(T,U)+\Xi(T,U)] \,,
\ee
where $\Xi$ is a quadratic polynomial in the variables
$(1,iT,iU,TU)$. Hence $\partial_T^3\Xi=\partial_U^3\Xi=0$.
The appearance of $\Xi$ complicates the symmetry properties of
the one-loop term, which would otherwise be a 
modular function of weight $-2$. However, the third derivative 
$\partial_T^3{\cal F}_{\rm 1-loop}$ transforms as a modular 
function of weight $+4$ under $T$-duality and of weight $-2$ under 
$U$-duality\footnote{%
   We note that, when $F(z') = (icz+d)^\nu \,G(z)$ with $z'\equiv 
   (az-ib)/(icz+d)$ and $ad-bc=1$, we have the following relation 
   for multiple derivatives:
   $$
   (\partial^nF)(z') = \sum_{k=0}^n \;\pmatrix{n\cr k\cr} 
   {\Gamma(\nu+n)\over\Gamma(\nu+k)} \,(ic)^{n-k} \,(icz+d)^{\nu +n+k} 
   \, \partial^kG(z)\, .
   $$
   When $n=1-\nu$, only the highest derivative survives on the 
   right-hand side.}.  %
The same statement applies to $\partial_U^3{\cal F}_{\rm 1-loop}$ 
with the modular weigths interchanged. The above result was also 
derived in \cite{A} for the specific case of the toroidal 
compactification. 

The polynomial $\Xi$ encodes the 
monodromies at singular points in the moduli space (for instance, 
at $T\approx U$) where one has an enhancement of the gauge symmetry. 
Knowledge of these singularities and of the 
asymptotic behaviour when $T\to \infty$ or $U\to \infty$, allows one to  
uniquely determine 
\be
\del_T^3 {\cal F}_{\rm 1-loop}(T,U)= {1\over2\pi}\,{E_4(iT)\, 
E_4(iU) \,E_6(iU) \over \big(j(iT)\,-\,j(iU)\big ) \,\eta^{24}(iU)}\,,
\ee
where $\eta$ is Dedekind's eta-function, $E_4$ and $E_6$ are the 
normalized Eisenstein's modular forms of respective weights $+4$ 
and $+6$ and $j$ is the modular invariant function 
$j=E_4^3/\eta^{24}$. A similar formula can be obtained for the 
third derivative with respect to $U$.  

We refer to \cite{DWKLL} for further details. Before closing we 
wish to point out that the dilaton field $S$ is no 
longer invariant under target-space duality in the presence of 
the one-loop corrections. This can be 
understood from the fact that the dilaton belongs originally to a 
vector-tensor multiplet and is only on-shell equivalent   
to a vector multiplet. However, one can 
always redefine $S$ such that it becomes invariant, 
but then it can no longer be interpreted as the scalar component 
of a vector multiplet \cite{DWKLL}. Interestingly enough, these 
perturbative results are confirmed by explicit calculations based on 
`string duality' between heterotic string compactifications on 
$K_3\times T_2$ and type-II string compactifications on  
Calabi-Yau manifolds \cite{stringdual}.   
 
\newpage

\noindent
This talk is based on work done in collaboration with V. 
Kaplunovsky, J. Louis, D. L\"ust and A. Van Proeyen. I thank B. 
Kleijn for useful comments on the manuscript.  \\[1mm]
This work was carried out in the framework of the European 
Community Research Programme ``Gauge theories, applied 
supersymmetry and quantum gravity", with a financial contribution 
under contract SC1-CT92-0789.

\end{document}